\documentclass{article}

\PassOptionsToPackage{numbers, compress}{natbib}

\usepackage[final]{neurips_2022}
\usepackage{natbib}
\usepackage[utf8]{inputenc} 
\usepackage[T1]{fontenc}    
\usepackage{hyperref}       
\usepackage{url}            
\usepackage{booktabs}       
\usepackage{amsfonts}       
\usepackage{nicefrac}       
\usepackage{microtype}      
\usepackage{xcolor}         

\usepackage{graphicx}
\usepackage{caption}
\usepackage{subcaption}
\usepackage{amsthm}
\newtheorem{definition}{Definition}

\title{Participation Interfaces for Human-Centered AI}

\author{%
  Sean McGregor \\ 
  IBM Watson AI XPRIZE and Responsible AI Collaborative\\
  Culver City, CA 90230 \\
  \texttt{hcai.interfaces@seanbmcgregor.com} \\
}

\begin{document}

\raggedbottom
\maketitle

\begin{abstract}
Emerging artificial intelligence (AI) applications often balance the preferences and impacts among diverse and contentious stakeholder groups. Accommodating these stakeholder groups during system design, development, and deployment requires tools for the elicitation of disparate system interests and collaboration interfaces supporting negotiation balancing those interests. This paper introduces interactive visual ``participation interfaces'' for Markov Decision Processes (MDPs) and collaborative ranking problems as examples restoring a human-centered locus of control.


\end{abstract}

Human-centered design has long been a software design philosophy centered on users \cite{NISTHCD,ISO13586}. In extending this practice to human-centered artificial intelligence (HCAI), the term ``human'' sometimes refers to \textit{users}, but increasingly often the term refers to \textit{humanity} writ large. As a matter of expediency, practitioners of HCAI distil the collective design target from ``humanity'' to that of the ``stakeholders'' affected by the intelligent system in question. Consistent with HCAI principles, the AI system should then ``empower them by making design choices that give humans control over technology'' (Shneiderman \cite{shneiderman2022human}). While trustworthy and responsible AI practices help identify and empower stakeholders in the design of AI systems, the technical question of how to map multiple competing stakeholder objectives to singular design choices is neither a solved problem nor one that is well supported by current technologies \cite{himmelreich2022against}.

We begin with the example of forest fire management, where there are multiple competing objectives advanced by different stakeholders. In a simulator-defined optimization of wildfire suppression policy, changing stakeholders between home owners and naturalists switches policy recommendations between suppressing all fires and letting all fires burn \cite{McGregor2019}. Should a forest management system then center on the stakeholder valuing ecology (the naturalist) or the stakeholder valuing clean air (the home owner)? These political questions do not have objective answers. Political questions require political processes. As a multi-discipline community, HCAI is well positioned to formalize political processes through the research of \textit{participation interfaces}.

\begin{definition}[HCAI Participation Interfaces]
  Systems by which people can understand stakeholder interests in an AI system and/or negotiate the trade-off space among those interests.
\end{definition}

While several research projects are examples of participation interfaces, such as WeBuildAI \cite{lee2019webuildai}, {\sc MDPv\/is} \cite{McGregor2019}, and JointReview (introduced below), supporting diverse stakeholder interests is not a current focus of HCAI \cite{xu2022transitioning}. We recommend the research and development of two types of HCAI interfaces.

\begin{figure}
     \centering
     \begin{subfigure}[b]{0.49\textwidth}
         \centering
         \includegraphics[width=\textwidth]{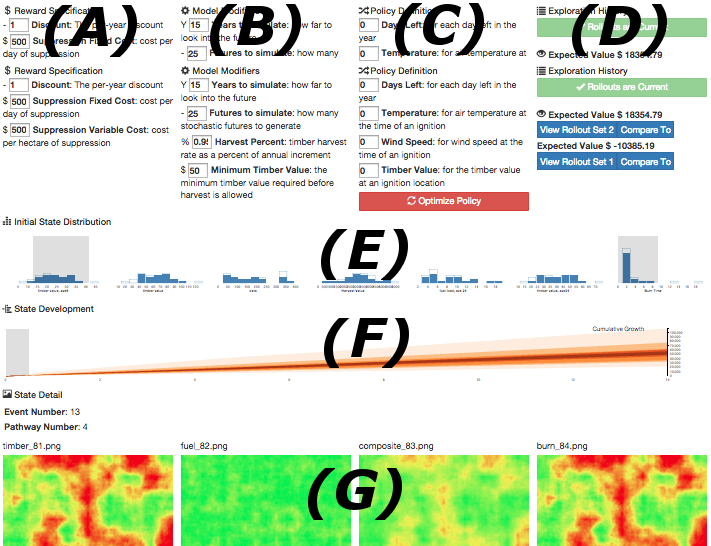}
         \caption{Markov Decision Process visualization, {\sc MDPv\/is}.
         The top row has
    parameter controls for (A) the reward
    function, (B) the model modifiers, (C) the policy
    definition, and (D) the history of configurations to (A) through
    (C). Changing parameters produces new optimizations subject to the parameters. The resulting outcomes are then rendered in panels for the initial state distribution (E), distributions through time (F), and images of the states (G). Complete details are in \citeauthor{McGregor2019} \cite{McGregor2019}}
         \label{fig:mdpvis}
     \end{subfigure}
     \hfill
     \begin{subfigure}[b]{0.49\textwidth}
         \centering
         \includegraphics[width=\textwidth]{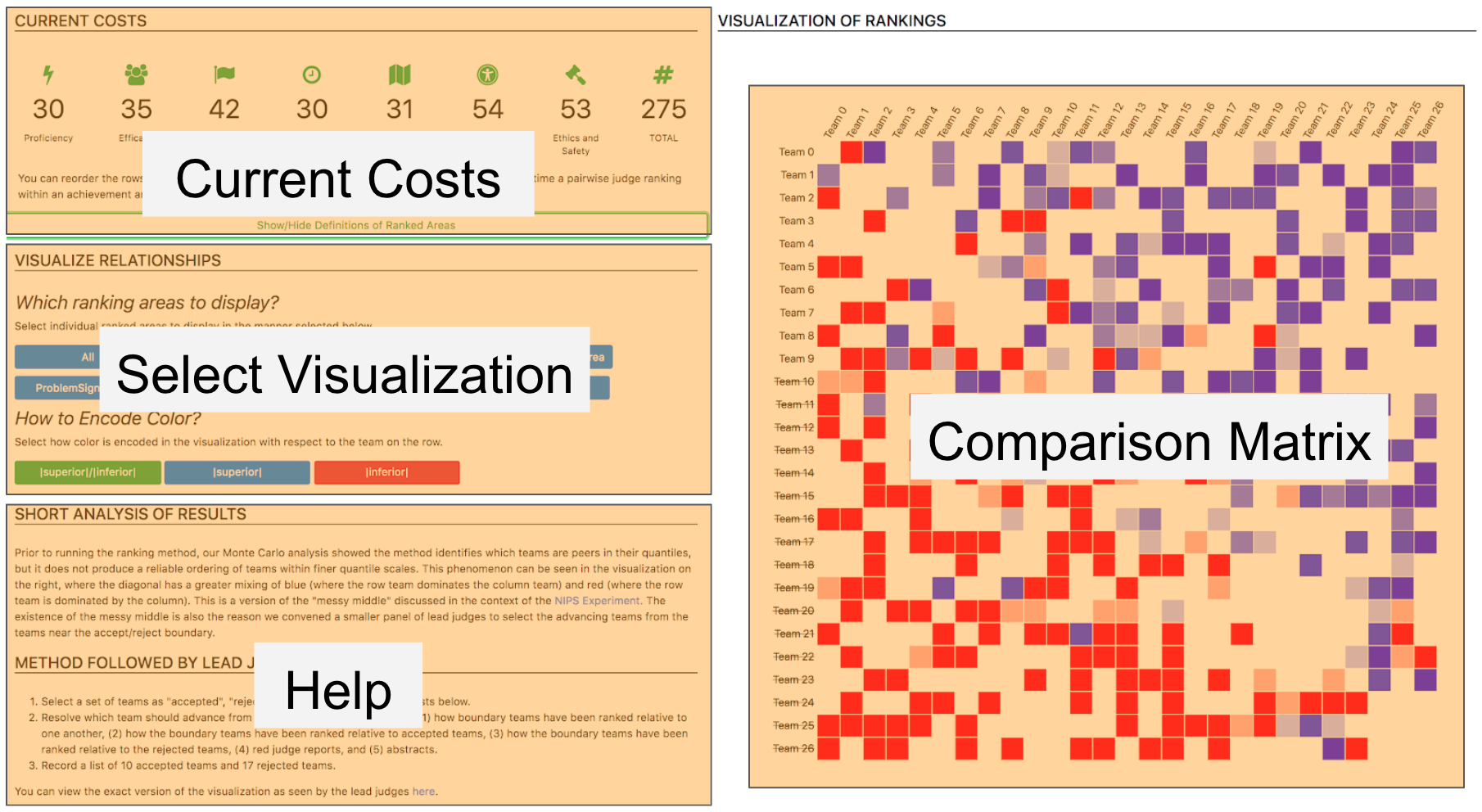}
         \caption{JointReview Visual analytic environment generated from reviewer force ranking of 27 submissions in the IBM Watson AI XPRIZE. The cost area presents a collection of scores where high scores indicate the current rank order in the comparison matrix violates a reviewer-ranked attribute. The review committee drags the rows of the comparison matrix when negotiating rankings and the costs recalculate. While a small subpanel of judges manipulated JointReview to produce the final decision, the optimization and costs represent the views of 30+ reviewing judges. Humans manipulating the interface can then negotiate simultaneously with each other, the machine's summaries, and the humans who informed the machine summaries.}
         \label{fig:jointreview}
     \end{subfigure}
     \hfill

\end{figure}

\textbf{(1) Interest Elicitation Interfaces.} Among the frameworks for formal negotiations is ``interest based bargaining'' (IBB), which seeks to form agreements on the basis of mutual and individual interests rather than positions \cite{FMCS}. IBB has been applied successfully in contentious labor-management contract negotiations \cite{CGE} to arrive at mutualistic outcomes. In the context of MDPs (i.e., reinforcement learning), interests are specified as ``reward functions'' that map outcomes to scalar benefit realized through time. Most stakeholders do not know the functional representation of their interests, so IBB can only successfully be applied after first solving the interest elicitation problem. We propose participation interfaces into models as the human-centered means of eliciting stakeholder interests. Figure \ref{fig:mdpvis} gives one example whereby interactive specification of the reward function and subsequent re-optimization provides interest insights to stakeholders.

\textbf{(2) Interest Negotiation Interfaces.} When interests are known, interfaces are then required to negotiate compromises. Taking the example of a particularly high-stakes instance of peer review, judges in the IBM Watson AI XPRIZE awarded a \$5 million dollar prize purse to teams producing the greatest advancements ``for good'' \cite{McGregor2018,AIMag2018,bondi2021envisioning}. High stakes ranking according to the ``good'' achieved presents a bounty of philosophical disagreements requiring negotiation. On the recommendations of \citeauthor{Shah2018} \cite{Shah2018} to collect rankings from each reviewer, the judges populated a sparse tensor of rankings that fed into a cost metric for machine optimization. The optimization produced a ``best'' rank ordering, but it would not be human-centered to simply take the recommendation of the machine. A subset of judges manipulated the suggested ranking as shown in Figure \ref{fig:jointreview} to make decisions. Each change to the ranking provided feedback from the machine's metrics, which the judges could weigh in their decisions. The judges ultimately chose to advance most -- but not all -- the teams selected by the algorithm. The combination of human inputs, machine optimization, and human negotiation forms a human-centered process for negotiating interests in collaborative ranking.

These two participation interfaces are among the many needed to ensure systems affecting more than one person are human-centered. While the field of explainable AI (XAI) provides many tools and techniques in the construction of these interfaces, the information flow is typically from the AI system to humans \cite{adadi2018peeking}, rather than a feedback loop between multiple humans and the intelligent system. Making effective use of machine recommendations requires systems where both the machine and humans can participate in the decision making process.


Participation interfaces are likely to be controversial when applied to decisions with profound differences among stakeholder outcomes \cite{sloane2020participation,chan2021limits,asaro2000transforming}. However, without participation interfaces the status quo ante will persist. Forests will continue to see ecological degradation, explosive subsequent fire seasons, and timber losses unbalanced by insights derived from AI. HCAI systems centered on facilitating participation processes can do better for wildfire problems, academic review, and any multi-stakeholder problem of divergent interests. AI adoption requires effective governance and so we must augment HCAI with systems for governance. 

Although participation interfaces do not inherently solve the problem of who hold the power to decide \cite{quick2011distinguishing,birhane2022power}, they do provide the capacity to more directly accommodate the interests of all humans.

\pagebreak

\textbf{ACKNOWLEDGMENT}

Many XPRIZE staff members and consultants supported the development and application of JointReview, including Amir Banifatemi, Neama Dadkhahnikoo, Devin Krotman, Jim Mainard, Daniel Miller, Roopa Dandamudi, Michael Martin, and Per-Erik Milam, among others.

Judges covering a wide array of disciplines participated in the IBM Watson AI XPRIZE review process over the course of the 4 year competition. They include, 
Gabriel Skantze, 
Eric Van Gieson, 
Adam Cheyer, 
Robin Murphy, 
Ivan Laptev, 
Alex London, 
Al Kellner, 
Erin Walker, 
Madeleine Clare Elish, 
Sidney D'Mello, 
Anna Bethke,
Julien Cornebise,
David Luxton,
David Kale, 
David Danks,
Danielle Tarraf, 
Xiaoyang Wang, 
Evan Muse, 
Nicolas Papernot, 
Risto Miikkulainen, 
Pascal Van Hentenryck, 
Mark Crowley, 
Forent Perronnin, 
Bill Smart, 
Graham Taylor, 
Becky Inkster,  
Julien Mairal, 
Stefano Ermon, 
Antoine Bordes, 
Jonathan Zittrain, 
Michael Gillam, 
Peter Eckersley, 
Casey Caruso,
Leila Toplic,
Rayid Ghani,
danah boyd,
and Barry O'Sullivan.

Next, the ``Red Judges'' performed the challenging task of simultaneously mentoring teams and representing the interests of society in general during the review processes that informed the presiding judges. They include, among others,
Jungo Kasai,
Duncan McElfresh,
Oliver Bent,
Ashis Pati,
Ines Nolasco,
Mayank Kejriwal,
Zeke Victor,
Marzieh Amini,
Sendong (Stan) Zhao,
Shahrzad Gholami,
Sabina Tomkins,
Joe Futoma,
Madian Khabsa,
Roy Fox,
Muhammed Ali Sit,
Marija Slavkovik,
Garrett Bernstein,
Elle Yuan Wang,
Randi Williams,
Tristan Sylvain,
Hoa Dam,
Safinah Ali,
Nicola Paoletti,
Kris Sankaran,
Kevin Winner,
Margaux Luck,
Tim Booher,
Colin Bellinger,
Heloïse Greeff,
Colin Bellinger,
Nour El Mawas,
Enno Hermann,
Dong Nie,
and
Muhammad Raza Khan.

Finally, the IBM Watson AI XPRIZE relied on an advisory board including Yoshua Bengio, Francesca Rossi, Rob High, Babak Hodjat, Neil Jacobstein, Subbarao (Rao) Kambhampati, Peter Norvig, Tim O'Reilly, Jean Ponce, Lav Varshney, and Manuela M. Veloso.

\bibliography{references}

\end{document}